# Scaling Law for the Cosmological Constant from Quantum Cosmology with Seven Extra Dimensions

Scott Funkhouser
Dept. of Physics, The Citadel, 171 Moultrie St. Charleston, SC 29409

ABSTRACT
According to a model of quantum cosmology the maximum number of degrees of freedom allowed in our three dimensions was determined by the size of seven extra dimensions in an initial excited state before inflation. The size of the extra dimensions can be inferred from a simple scheme for unifying the strong force and gravity. Coupled with the Bekenstein-Hawking entropy bound, these considerations lead to a scaling law for the cosmological constant that has been proposed independently by several authors.

1. Introduction

In a simple quantum cosmology the Universe began with a quantum fluctuation from nothing into a homogeneous 10-dimensional space. That quantum state then tunneled into another state in which seven of the dimensions were excited, subsequently collapsing to their ground state. Certain parameters of the excited and ground states of the putative extra dimensions can be specified by considering the relative strengths of the gravitational and strong forces. In order to unify gravity and the nuclear force in the very early universe the gravitational force must have been much stronger at the beginning of inflation and it must have weakened during inflation. That change in the strength of the gravitational force may be modeled by allowing the Newtonian gravitational coupling $G$ to vary. The size of the extra dimensions would correspond to the characteristic microscopic length scale associated with the gravitational coupling, and would decrease as gravity weakened. (Mongan 2001)

In a simple unification scheme, the initial gravitational coupling $G_i$ would have been such that

$$\frac{G_i m_n^2}{\hbar c} = 1, \qquad (1)$$

where $m_n$ is the nucleon mass, $\hbar$ is the Planck constant and $c$ is the speed of light in a vacuum. Eq. (1) leads directly to

$$\frac{G_i}{G} = \frac{m_P^2}{m_n^2} \approx 1.7 \times 10^{38}, \qquad (2)$$

where $m_P$ is the Planck mass. As the seven compact dimensions collapsed from the excited state to the ground state, the gravitational coupling $G_i$ must have decreased to $G$. In the excited state the size of the extra dimensions would have been roughly $l_i \equiv (\hbar G_i / c^3)^{1/2}$, which is the microscopic length associated with $G_i$. During inflation, as the gravitational coupling decreased to $G$, the size of the extra dimensions decreased to the Planck length $l_P$. Note that the size of the extra dimensions $l_i$ in the excited state would have been, by Eq. (2), roughly equal to the Compton wavelength of the nucleon $l_n$. (Mongan 2001)

In that scenario the available degrees of freedom associated with the seven extra dimensions were transferred to the three-dimensional universe during inflation. According to the Bekenstein-Hawking entropy bound, the maximum number of degrees of freedom allowed in a certain space is one fourth of the surface area of the space measured in Planck units (Hawking 1975). The surface area $A_7(r)$ of a seven-dimensional sphere of radius $r$ is given by

$$A_7(r) = \frac{16\pi^3}{15} r^6. \tag{3}$$

The actual area $A_7' \equiv fA_7$ of the seven-dimensional space that was part of the early Universe in its excited state may have been larger than the area in Eq. (3) by a positive, real factor $f$ that represents possible convolutions of the surfaces. It is presumed here that $f$ is of order near $10^0$. If the characteristic radii of the extra dimensions were initially of order the nucleon's Compton wavelength $l_n$, the maximum number of degrees of freedom allowed in the initial seven-dimensional subspace would have been roughly

$$N_7 \sim \frac{A_7'(l_i)}{l_P^6} \sim f^6 \left(\frac{l_n}{l_P}\right)^6, \tag{4}$$

which is of the order $10^{119} f^6$. (Mongan 2003)

2. A Scaling Law for the Cosmological Constant

According to the model outlined by Mongan, $N_7$ should be also the maximum number of degrees of freedom that could be contained in our three dimensions (Mongan 2003). Not only does that prediction agree with calculations of the maximum entropy of our Universe, but it also leads to a known scaling law for the cosmological constant. The maximum number of degrees of freedom that could ever be contained in our three dimensions is determined by the cosmological constant $\Lambda$. The de Sitter horizon $R_\Lambda$ is the largest possible event horizon in a vacuum-dominated Universe, and it is equal to $c\sqrt{3/\Lambda}$. According to the Bekenstein-Hawking entropy bound, the maximum number $N_\Lambda$ of degrees of freedom allowed in a sphere whose radius is the de Sitter horizon is

$$N_\Lambda = \frac{\pi R_\Lambda^2}{l_P^2} \sim \frac{c^5}{G\hbar\Lambda}. \tag{5}$$

If the current Hubble parameter is 71km/s/Mpc and if the fraction of the cosmic energy inventory due to the vacuum is 0.73 then the cosmological constant is roughly $1.2\times10^{-35}$s$^{-2}$. With that value for $\Lambda$, the number $N_\Lambda$ is roughly $2.5\times10^{122}$.

Given the uncertainty of a few orders of magnitude associated with the factor $f$ and other unknown coefficients, the maximum number of degrees of freedom of our universe $N_\Lambda \sim 10^{122}$ agrees well with the number $N_7 \sim 10^{119}$ that follows from basic considerations of a simple quantum cosmological model. Furthermore, equating $N_\Lambda$ to $N_7$ leads to a scaling relationship between the cosmological constant and other fundamental parameters of nature,

$$\Lambda \sim \frac{G^2 c^2 m_n^6}{\hbar^4}. \tag{6}$$

That is to say that the cosmological constant would be scaled to the sixth power of the nucleon mass, or *vice versa*. Remarkably, the scaling in Eq. (6) is identical to the scaling law proposed independently by Zel'dovich (Zel'dovich 1967), Mena Marugan and Carneiro (Mena 2002), and this author (Funkhouser 2006). The fact that the model in Ref. (Mongan 2003) leads to a good prediction for the maximum entropy of our universe and a known scaling law for the cosmological constant is a compelling motivation to consider that both Eq. (6) and quantum cosmology with seven extra dimensions are physically meaningful.

Acknowledgments: This work benefited from conversations with T. Mongan.